\documentclass[10pt]{article}
\usepackage{graphicx}

\def\Title#1{\begin{center} {\Large #1 } \end{center}}

\def\Author#1{\begin{center}{ \sc #1} \end{center}}

\def\Address#1{\begin{center}{ \it #1} \end{center}}

\newcommand\pubblock{\rightline{\begin{tabular}{l} Proceedings of the Second Annual LHCP\\ \pubnumber\\
         \pubdate  \end{tabular}}}

\newenvironment{Abstract}{\begin{quotation} 
\begin{center} 
             
\large ABSTRACT \end{center}\bigskip 
     
 \begin{center}\begin{large}}
{\end{large}\end{center} 
\end{quotation}}

\newenvironment{Presented}{\begin{quotation} \begin{center} 
             PRESENTED AT\end{center}\bigskip 
      \begin{center}\begin{large}}{\end{large}\end{center} \end{quotation}}

\def\Acknowledgements{\bigskip  \bigskip \begin{center} \begin{large}
             \bf ACKNOWLEDGEMENTS \end{large}\end{center}}

\def\beq{\begin{equation}}
\def\eeq#1{\label{#1}\end{equation}}
\def\eeqn{\end{equation}}

\def\beqa{\begin{eqnarray}}
\def\eeqa#1{\label{#1}\end{eqnarray}}
\def\eeqan{\end{eqnarray}}

\def\overbar#1{\overline{#1}}

\let\bar=\overbar

\def\Dslash{\not{\hbox{\kern-4pt $D$}}}
\def\dslash{\not{\hbox{\kern-2pt $\del$}}}

\def\msb{{\bar{\ssstyle M \kern -1pt S}}}

\usepackage{color}
\usepackage{float} 

\newcommand\pubnumber{ FERMILAB-CONF-14-262-E }

\newcommand\pubdate{September 1, 2014}

\def\affiliation{
On behalf of the CDF and D0 Collaborations
\\
Fermi National Accelerator Laboratory 
\\
Batavia, IL 60510, U.S.A.}

\begin{document}



\large
\begin{titlepage}
\pubblock



\vfill
\Title{  RECENT TEVATRON RESULTS ON $CP$-VIOLATION  }

\vfill



\Author{ Peter H. Garbincius  }

\Address{\affiliation}
\vfill

\begin{Abstract}

Using 
their full Tevatron Run II data sets, 
the CDF and D0 Experiments present measurements of $CP$-violating asymmetries 
in the charmless decays of bottom baryons
 $\Lambda_b^0 \rightarrow  p \pi^-$, $\Lambda_b^0 \rightarrow p K^-$, and also for
 $B_s^0 \rightarrow K^- \pi^+$, $B^0 \rightarrow K^+ \pi^-$, $D_s^\pm \rightarrow \phi \pi^\pm$,
 and for single muons and like-sign dimuons in $p\overline{p}$ collisions. 
 Except for the like-sign dimuon asymmetry,
 these asymmetry measurements are consistent with available predictions of the standard model.

\end{Abstract}
\vfill


\begin{Presented}
The Second Annual Conference
\\
 on Large Hadron Collider Physics 
\\
Columbia University, New York, U.S.A.
\\ 
June 2-7, 2014

\end{Presented}

\vfill

\end{titlepage}

\def
\thefootnote{\fnsymbol{footnote}}

\setcounter{footnote}{0}

%


\normalsize




















\section{Introduction}

One of the great, outstanding mysteries of nature is that we observe 
a universe consisting of matter, rather than a very different type of universe consisting 
of equal quantities of matter and antimatter.  The violation of Charge-Parity ($CP$) symmetry 
is a necessary ingredient \cite{Sakharov} in producing such a matter-dominated universe.  
So far, small amounts of $CP$-violation have been observed only in the neutral kaon and 
neutral 
$b$-meson systems.  $CP$-violation is included in the standard model (SM), which reproduces 
almost all observed particle physics processes. 
By searching for and measuring $CP$-violating 
processes and comparing with the SM predictions, we may open a window to observe additional $CP$-violation 
produced by other processes, or new physics, beyond the standard model.

Final results for the $CP$-violating asymmetries are presented for the full data sets for the CDF 
(9.3 fb$^{-1}$) and D0 (10.4 fb$^{-1}$) experiments.  The Fermilab Tevatron produced $p\overline{p}$ collisions 
at $\sqrt{s}$ = 1.96 TeV until its shutdown at the end of September, 2011.  $p\overline{p}$ collisions are 
ideal for $CP$ studies since the initial state is symmetrical between particle and anti-particles.  
The $e^+e^-$ $b$-factories do not have the kinematic reach at reasonable luminosity to study the $B_s^0$ system.  
Both the CDF  II \cite{CDF description} and D0 \cite{D0 description} detectors are symmetric in 
$\eta$~=~-ln(tan($\theta$/2)) where $\theta$ is the polar angle of the produced particles relative to the proton beam 
direction, which gives equal acceptances even if there is a forward-backward production asymmetry for the 
detected final state.  Both CDF and D0 detectors have silicon vertex detectors to study displaced 
vertices from decays of particles containing $b$- or $c$-quarks.  In addition, the CDF II detector 
features a displaced vertex trigger and particle identification by energy loss $dE/dx$ and time of 
flight $TOF$ in the central drift chamber. CDF also can separate $\Lambda_b^0 \rightarrow
pK^-$ from $\overline{\Lambda_b^0} \rightarrow \overline{p} K^+$
by the kinematic momentum asymmetry $(p_+ - p_-)/(p_+ + p_-)$ between the positively and negatively charged particles in the decay.   
D0 features excellent 
muon identification using the muon toroid magnets and muon chambers and scintillators.  D0 also 
regularly flips the polarities of the solenoid and toroid magnets independently, giving approximately 
equal integrated luminosities for all four configurations, which cancels many of the systematic, 
detector acceptance asymmetries. 

The $CP$ asymmetries are defined as the ratio of the difference divided by the sum of the rates 
$\Gamma($particle$ \rightarrow f)$ and $\Gamma($antiparticle$ \rightarrow \overline{f})$ for processes involving particles and anti-particles: 
\begin{center}
$A_{CP} = [\Gamma(b \rightarrow f) - \Gamma(\overline{b} \rightarrow \overline{f})]/[\Gamma(b \rightarrow f) + \Gamma(\overline{b} \rightarrow \overline{f})]
 = [N_{b \rightarrow f} - c_f*N_{\overline{b} \rightarrow \overline{f}}] / [N_{b \rightarrow f} + c_f*N_{\overline{b} \rightarrow \overline{f}}]$
\end{center}
where $f(\overline{f})$ is the final state in the particle(antiparticle) decay.
The raw number of observed particle and antiparticle events are corrected for detector 
efficiency ratio  $c_f = \varepsilon(f)/\varepsilon(\overline{f})$ and detector related asymmetries to calculate $A_{CP}$.




\section{CDF: $A_{CP}$ in Charmless Decays of Bottom Hadrons }

\begin{figure}[htb]

\centering

\includegraphics[height=2in]{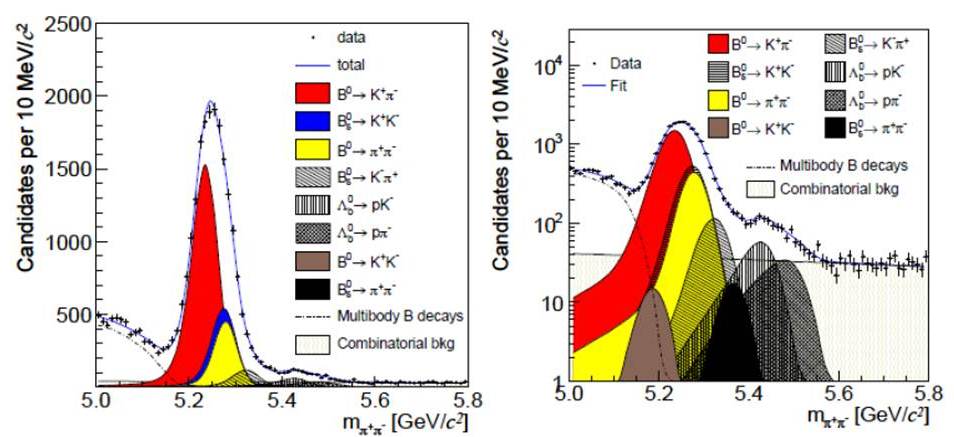}

\caption{ CDF fits to the two-body invariant mass distribution where the two final state 
particles are both assigned the pion mass.  This shifts the mass peaks depending which particle 
is really a kaon or proton (antiproton) and on the mass of the parent particle, allowing the 
simultaneous fitting of the number of individual decay processes.}

\label{fig:CDF}
\end{figure}

This study is motivated by possible new physics from internal penguin loops.  The SM expects that
$A_{CP}(B^0 \rightarrow K^+ \pi^-)$ would equal $A_{CP}(B^+ \rightarrow K^+ \pi^0)$.  However, $b$-factory measurements \cite{PDGavg,Belle} find 
that these $A_{CP}$s are different by a 4-5 $\sigma$ discrepancy.  This $\Delta A_{B \rightarrow K \pi}$ puzzle may indicate incomplete understanding within the SM.
Figure \ref{fig:CDF} shows the CDF \cite{CDF} data fits. The results are presented in Table \ref{tab:table1} and discussed below.




%


%





CDF finds a significantly non-zero $A_{CP}(B^0 \rightarrow  K^+ \pi^-)$ consistent with the measurements of LHCb \cite{LHCb}, 
BaBar \cite{BaBar3}, and Belle \cite{Belle}.


Similarly CDF observes a significantly non-zero $A_{CP}(B_s^0 \rightarrow K^- \pi^+) = [22 \pm 7 $(stat.)$ \pm 2 $(syst.)$]\%$, which is consistent 
with the LHCb measurement \cite{LHCb}, and as expected by U-spin symmetry 
\cite{isospin}, is consistent with the direct, non-oscillated $A_{CP}^{direct}(B^0 \rightarrow \pi^+ \pi^-)$ \cite{Belle pi pi}. 
CDF's $A_{CP}(B_s^0 \rightarrow K^- \pi^+)$ measurement is also consistent with the expected value
by Gronau-Rosner-Lipkin scaling \cite{Gronau-Lipkin} according to the SM
\begin{center}
$A_{CP}(B_s^0 \rightarrow K^- \pi^+) = -A_{CP}(B^0 \rightarrow K^+ \pi^-)*{\cal B}(B^0 \rightarrow K^+ \pi^-)/{\cal B}(B_s^0 \rightarrow K^- \pi^+)*\tau(B_s^0)/\tau(B^0)$ .
\end{center}

The unique CDF \cite{CDF} observations $A_{CP}(\Lambda_b^0 \rightarrow p\pi^-) = [+6 \pm 7 $(stat.)$ \pm 3 $(syst.)$]\%$ and $A_{CP}(\Lambda_b^0 \rightarrow p K^-) = 
[-10 \pm 8 $(stat.)$ \pm 4 $(syst.)$]\%$ are both consistent with zero, excluding large asymmetries.  There are no SM calculations of comparable precision 
with which to compare.

%
\begin{table}[t]
\begin{center}
\begin{tabular}{ll}  
\hline
\hline
$A_{CP}(B^0 \rightarrow K^+ \pi^-)$ & SM predictions complicated by hadronic decays \\
 $[-8.3 \pm 1.3 $(stat.)$ \pm 0.4 $(syst.)$]\%$ &  CDF (2014) \cite{CDF}\\
 $[-8.0 \pm 0.7 $(stat.)$ \pm 0.3 $(syst.)$]\%$ & LHCb (2013) \cite{LHCb} \\
 $[-10.7 \pm 1.6 $(stat.)$ \pm 0.5 $(syst.)$]\%$ & BaBar (2013) \cite{BaBar3} \\
 $[-6.9 \pm 1.4 $(stat.)$ \pm 0.7 $(syst.)$]\%$ &  Belle (2013) \cite{Belle} \\
\hline
$A_{CP}(B_s^0 \rightarrow K^- \pi^+)$ & \\
$[22 \pm 7 $(stat.)$ \pm 2 $(syst.)$]\%$ &  CDF (2014) \cite{CDF} \\
$[27 \pm 4 $(stat.)$ \pm 1 $(syst.)$]\%$ & LHCb (2013) \cite{LHCb} \\
$[33 \pm 6 $(stat.)$ \pm 3 $(syst.)$]\%$ & Belle (2013) \cite{Belle pi pi} $A_{CP}^{direct}(B^0 \rightarrow \pi^+ \pi^-)$ by U-spin symmetry \cite{isospin} \\
$[31 \pm 5 $(stat.+syst.)$]\%$ & SM scaling \cite{Gronau-Lipkin} $A_{CP}(B^0 \rightarrow K^+ \pi^-)$ using PDG data \cite{PDGavg} \\
\hline
\hline
\end{tabular}
\caption{$A_{CP}(B^0 \rightarrow K^+ \pi^-)$ and $A_{CP}(B_s^0 \rightarrow K^- \pi^+)$ for CDF and other experiments, by U-spin symmetry, and by  Gronau-Rosner-Lipkin scaling.   }
\label{tab:table1}
\end{center}
\end{table}



\section{D0:  $A_{CP}(D_s^\pm \rightarrow \phi\pi^\pm)$ }

\begin{figure}[htb]

\centering

\includegraphics[height=2.2in]{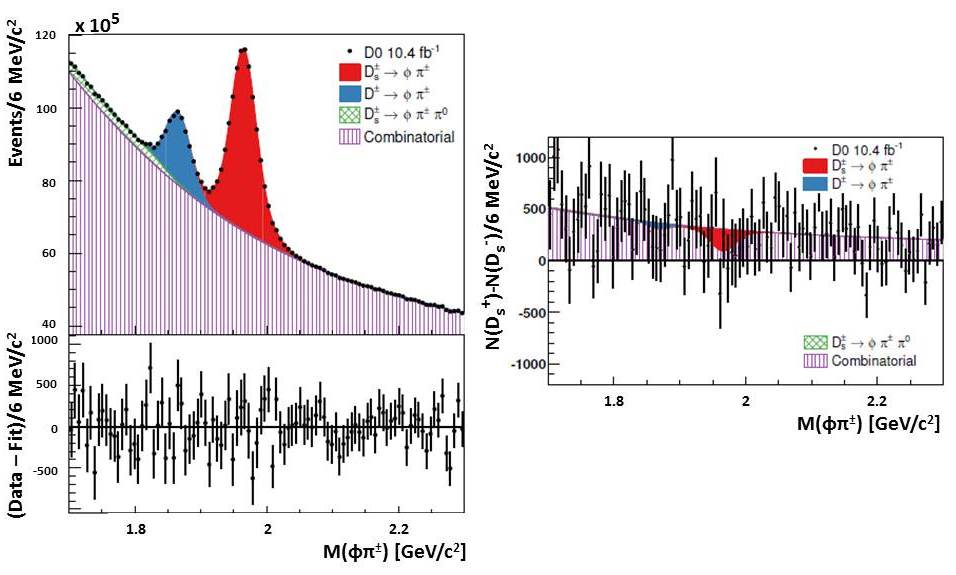}

\caption{ The $\phi\pi^\pm$ invariant mass distributions for the D0 Experiment.  The upper-left panel shows
the sum of the positively and negatively charged candidate events.  
The lower mass peak (blue) is due to the $D^\pm$ decays while the higher mass peak (red) is due to the 
$D_s^\pm$ meson decays. The lower-left panel 
shows the fit residuals for the sum distribution above.  
The right panel shows the fit to the differences between the numbers of candidate $D_s^+$ and $D_s^-$ 
mesons as function of the $\phi\pi^\pm$ mass.}

\label{fig:Ds}
\end{figure}

The SM predicts zero $A_{CP}$ for the $D_s^\pm \rightarrow \phi\pi^\pm$ decay. 
By fitting the observed sum and difference of the number of observed $D_s^+$ and $D_s^-$ events in 
Figure \ref{fig:Ds}, D0 \cite{D0 Ds} finds $A_{raw}(D_s) = (-0.43 \pm 0.26)\%$.  
Correcting for the detector and background asymmetries $A_{KK}+A_\mu$ and the physics-driven $A_{physics}$
mainly due to the $B_s^0-\overline{B_s^0}$ oscillation and the subsequent semileptonic decay $B_s^0 \rightarrow \mu^+ \nu D_s^- X$  
\begin{center}
$A_{CP}(D_s^\pm \rightarrow \phi \pi^\pm) = A_{raw} -  A_{KK} - A_{\mu} - A_{physics} = [-0.38 \pm 0.26 ($stat.$) \pm 0.08 ($syst.$)]\%$, 
\end{center}
which is consistent with the SM prediction of zero and greater than 3 times more sensitive 
than the earlier CLEO limit \cite{CLEO} of $[-0.5 \pm 0.8 ($stat.$) \pm 0.4 ($syst.$)]\%$.

%


%



\section{D0:  Like-Sign Dimuon Charge Asymmetry }

The SM produces like-sign $\mu^\pm \mu^\pm$ pairs from $B^0-\overline{B^0}$ and $B_s^0-\overline{B_s^0}$ mixing followed 
by semileptonic decays.  $CP$-violating asymmetries would be produced if the rates 
of mixing differ $\Gamma(B^0 \rightarrow \overline{B^0}) \not= \Gamma(\overline{B^0} \rightarrow B^0)$ 
and/or $\Gamma(B_s^0 \rightarrow \overline{B_s^0}) \not= \Gamma(\overline{B_s^0} \rightarrow B_s^0)$.  
There is also a small $CP$-violating contribution from interference between mixed and 
direct decays to states accessible to both the particle and antiparticle parents such 
as $B^0$ or $\overline{B^0} \rightarrow D^{(*)+} + D^{(*)-}$ followed by $D^{(*)\pm} \rightarrow \mu^{\pm} X$ ~\cite{BandH}.

The prior three D0 measurements at 1 fb$^{-1}$ \cite{D0-1}, 6.1 fb$^{-1}$ \cite{D0-6}, and
9 fb$^{-1}$ \cite{D0-9} observed the like-sign dimuon charge asymmetry
	$A_{CP} = [\Gamma(\mu^+ \mu^+) -  \Gamma(\mu^- \mu^-)]/[\Gamma(\mu^+ \mu^+) + \Gamma(\mu^-\mu^-)]$ 
to be inconsistent with the prediction of the SM at 1.7-3.9 $\sigma$ significance.  
This D0 final measurement \cite{D0 2mu} includes the full 10.4 fb$^{-1}$ Run II data sample and improved 
methodology and background subtraction which includes tighter track quality requirements 
and $A_{CP}$ measurements for each of three bins in muon Impact Parameter (IP).
Corrections were applied for measured charge-dependent
asymmetries in efficiencies for detecting muons, and for pions and kaons faking 
muons.



\begin{figure}[htb]

\centering

\includegraphics[height=2.1in]{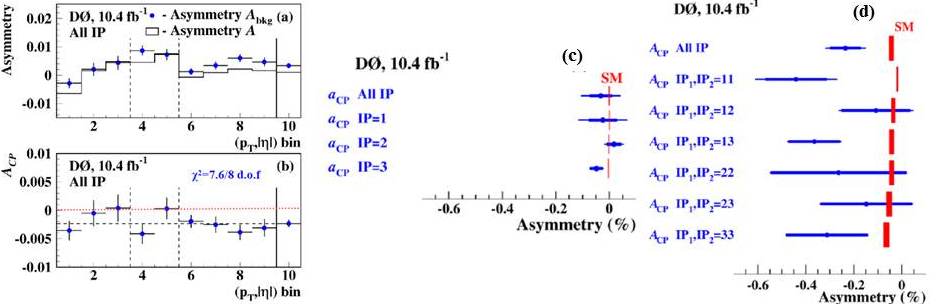}

\caption{ (a)  The histogram is the raw asymmetry $A$, the data points 
are the measured detector-introduced background $A_{bkg}$; 
(b) $A_{CP} = A - A_{bkd}$ 
with the average  $A_{CP}$ value = dotted horizontal line and $A_{CP}$ equal zero = red horizontal line.  
The bins 1-3 are for $|\eta| < 0.7$, bins 4-5 correspond to $0.7 < |\eta| < 1.2$, and bins 6-9 
correspond to $1.2 < |\eta| < 2.2$ ranges for the dimuon pairs.  The right-most bin $\#$10 is 
integrated over all 9 ($p_T,|\eta|$) bins and integrated over all IP;
(c) The single muon charge asymmetry $a_{CP}$ for different IP samples.  The dashed line represents the SM prediction;
(d) The like-sign dimuon asymmetry $A_{CP}$ for different IP$_1$,IP$_2$ samples. The boxes show the SM prediction with the box width 
corresponding to the theoretical uncertainty.}
\label{fig:twomu}
\end{figure}

 Integrating over all IP, $|\eta|$, $p_T$ bins, D0 measures the inclusive single muon charge asymmetry 
 $a_{CP}$
(all IP) = $[-0.032 \pm 0.042 ($stat.$) \pm 0.061 ($syst.$)]\%$ which is consistent with the SM prediction of 
$a_{CP} \sim 10^{-5}$. The raw asymmetry $a$ varies 
considerably and even changes sign over this range, giving confidence that no artificial 
asymmetry is introduced by the detector or analysis.
The measured like-sign dimuon charge asymmetry dependence on the ($p_T,|\eta|$) bins 
(summed over all IP) is shown in Figure \ref{fig:twomu}.  
The final observed like-sign dimuon charge asymmetry is measured to be
$A_{CP} = [-0.235 \pm 0.065 ($stat.$) \pm 0.054 ($syst.$)]\%$ which is consistent with 
the three prior D0 measurements and represents a 3.6 standard deviation discrepancy
significance from the SM prediction in Figure \ref{fig:twomu} and a 4.1 $\sigma$ deviation from $A_{CP} = 0$.


%







%




%


%



\begin{figure}[h!]

\centering

\includegraphics[height=2.5in]{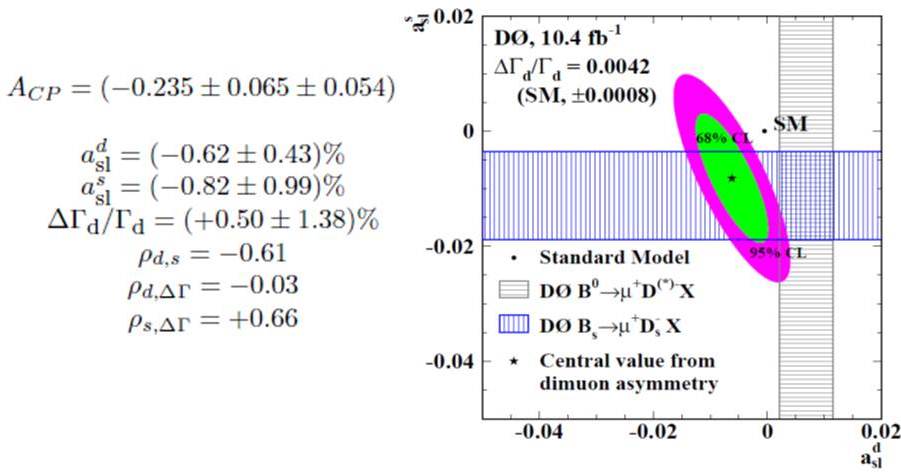}

\caption{The D0 fitted parameters, including correlations, for the like-sign dimuon asymmetry. The figure shows the 68\% and 95\% confidence level contours 
in the $a^d_\textup{\small{sl}} $ - $a^s_\textup{\small{sl}} $ plane obtained 
from the fit of the inclusive single muon and like-sign dimuon asymmetries with fixed 
value of 
$\Delta\Gamma_\textup{\footnotesize{d}}$/$\Gamma_\textup{\footnotesize{d}}$ corresponding to the expected SM value \cite{L-N}. Also shown are the prior D0 semileptonic 
measurements of $a^d_\textup{\small{sl}} $ \cite{D0-asld} and $a^s_\textup{\small{sl}} $ \cite{D0-asls}, with bands representing $\pm 1$ standard deviation uncertainties of these measurements.}

\label{fig:aslq}
\end{figure}

The like-sign dimuon charge asymmetry for each kinematic (IP, $p_T,|\eta|$) bin is a 
linear function of the semileptonic charge asymmetries for $B^0 \rightarrow \mu^+ D^- X$ and 
$B_s^0 \rightarrow \mu^+ D_s^- X$ mesons, and the decay rate difference between the light and heavy 
components of the $B^0-\overline{B^0}$ system
\begin{center}
$A_{CP} = C_d*a^d_\textup{\small{sl}}$ + $C_s*a^s_\textup{\small{sl}}$ + $ C_\delta*\Delta\Gamma_\textup{\footnotesize{d}}$/$\Gamma_\textup{\footnotesize{d}}$
\end{center}
where the linear coefficients $C_d$, $C_s$, and $C_\delta$ depend on the particular kinematic bin. These measured physics parameters are listed and plotted in Figure \ref{fig:aslq}, showing consistency with prior 
D0 measurements of the $a^d_\textup{\small{sl}} $ \cite{D0-asld} and $a^s_\textup{\small{sl}} $ \cite{D0-asls} semileptonic decay asymmetries.

%
%

%


%


%



\section{Summary}

The recent Tevatron measurements of direct $CP$-violating asymmetries in charmless 
decays of bottom baryons and mesons by CDF and for $D^\pm \rightarrow \phi \pi^\pm$ by D0 are consistent 
with available SM predictions. 
All experiments find $A_{CP}(B^0 \rightarrow K^+ \pi^-)$ to be quite significantly different from zero.
The final D0 measurement of the CP-violating charge 
asymmetry of like-sign dimuons in $p\overbar{p}$ collisions is still inconsistent with the SM 
prediction at the 3.6 standard deviation level, which represents one of the few such 
inconsistencies.

\Acknowledgements
The CDF Collaboration thanks the Fermilab staff and the technical staffs of the 
participating institutions for their vital contributions. This work was supported 
by the U.S. Department of Energy and National Science Foundation; the Italian
 Istituto Nazionale di Fisica Nucleare; the Ministry of Education, Culture, Sports, 
Science and Technology of Japan; the Natural Sciences and Engineering Research 
Council of Canada; the National Science Council of the Republic of China; the 
Swiss National Science Foundation;  the A.P. Sloan Foundation; the Bundesministerium 
f\"{u}r Bildung und Forschung, Germany; the Korean World Class University Program, the 
National Research Foundation of Korea; the Science and Technology Facilities Council 
and the Royal Society, United Kingdom; the Russian Foundation for Basic Research; 
the Ministerio de Ciencia e Innovaci\'{o}n, and Programa Consolider-Ingenio 2010, 
Spain; the Slovak R\&D Agency; the Academy of Finland; the Australian Research Council 
(ARC); and the EU community Marie Curie Fellowship Contract No. 302103.

The D0 Collaboration thanks the staffs at Fermilab and collaborating institutions, 
and acknowledge support from the DOE and NSF (USA); CEA and CNRS/IN2P3 (France); MON, 
NRC KI and RFBR (Russia); CNPq, FAPERJ, FAPESP and FUNDUNESP (Brazil); DAE and DST 
(India); Colciencias (Colombia); CONACyT (Mexico); NRF (Korea); FOM (The Netherlands); 
STFC and the Royal Society (United Kingdom); MSMT and GACR (Czech Republic); BMBF and DFG 
(Germany); SFI (Ireland); The Swedish Research Council (Sweden); and CAS and CNSF(China).

%

\end{document}